\documentclass[letterpaper]{article} 
\usepackage{aaai2026}  
\usepackage{times}  
\usepackage{helvet}  
\usepackage{courier}  
\usepackage[hyphens]{url}  
\usepackage{graphicx} 
\usepackage{natbib}  
\usepackage{caption} 
\usepackage{algorithm}
\usepackage{algorithmic}

\usepackage{booktabs}
\usepackage{adjustbox}
\usepackage{graphicx}
\usepackage{multirow}
\usepackage{array} 

\usepackage{newfloat}
\usepackage{listings}
\DeclareCaptionStyle{ruled}{labelfont=normalfont,labelsep=colon,strut=off} 
\floatstyle{ruled}
\newfloat{listing}{tb}{lst}{}
\floatname{listing}{Listing}
\title{Towards Better Search with Domain-Aware Text Embeddings for C2C Marketplaces}
\author {
    Andre Rusli,
    Miao Cao,
    Shoma Ishimoto,
    Sho Akiyama,
    Max Frenzel
}
\affiliations {
    Mercari, Inc.\\
    {\{andre.rusli, miao, s-a-ishimoto, s-akiyama, m-frenzel\}@mercari.com}
}

\usepackage{bibentry}

\begin{document}

\maketitle

\begin{abstract}
Consumer-to-consumer (C2C) marketplaces pose distinct retrieval challenges: short, ambiguous queries; noisy, user-generated listings; and strict production constraints. This paper reports our experiment to build a domain-aware Japanese text-embedding approach to improve the quality of search at Mercari, Japan's largest C2C marketplace. We experimented with fine-tuning on purchase-driven query-title pairs, using role-specific prefixes to model query–item asymmetry. To meet production constraints, we apply Matryoshka Representation Learning to obtain compact, truncation-robust embeddings. Offline evaluation on historical search logs shows consistent gains over a strong generic encoder, with particularly large improvements when replacing PCA compression with Matryoshka truncation. A manual assessment further highlights better handling of proper nouns, marketplace-specific semantics, and term-importance alignment. Additionally, an initial online A/B test demonstrates statistically significant improvements in revenue per user and search-flow efficiency, with transaction frequency maintained. Results show that domain-aware embeddings improve relevance and efficiency at scale and form a practical foundation for richer LLM-era search experiences.
\end{abstract}


\section{Introduction}

Search is a critical component of online consumer-to-consumer (C2C) marketplaces where listings are user-generated, diverse, and often noisy. Additionally, product search differs materially from web search \cite{nigam2019semanticproductsearch}. Queries are shorter, and strong relevance signals (purchases) are far sparser than clicks, which can cause click-driven models to over-rank accessories (e.g., cases, straps) over the core item (e.g., a phone or Nintendo Switch). Sessions often mix intents (exact model lookup, compatible accessories, substitutes, deal hunting), so systems must both minimize effort for mission-oriented queries and still support broad exploration. 

Dense, embedding-based matching has become a standard remedy for vocabulary mismatch and semantic variability in e-commerce search \cite{muhamed2023WebScaleSemanticProductSearch}. In parallel, general-purpose text-embedding models, trained on large and heterogeneous corpora, now deliver strong zero-/few-shot retrieval across tasks \cite{wang2024textembeddingsweaklysupervisedcontrastive,li2023generaltextembeddingsmultistage,gunther2024jinaembeddings28192token}. Recent progress extends beyond English via synthetic supervision from LLMs, enabling competitive Japanese text embeddings \cite{tsukagoshi2024rurijapanesegeneraltext}.

Yet, in C2C marketplaces, off-the-shelf Japanese text embeddings remain insufficient: short, ambiguous queries and user-generated listings demand domain-aware representations that encode product relations, and marketplace-specific semantics. For instance, a short query like ``switch'' may refer to the Nintendo Switch console, but can also surface unrelated results such as electrical switches or clothing items. Such ambiguity, combined with the diversity and pace of the marketplace ecosystem, frequently leads to noisy retrieval. Moreover, meeting latency/throughput at scale requires compact embeddings without sacrificing retrieval quality.

This paper reports our experiment to build a domain-aware Japanese text-embedding model for C2C search at Mercari, Japan’s largest C2C marketplace. Specifically, we fine-tune a text encoder using role-specific prefixes with Multiple Negatives Ranking (in-batch negatives) \cite{henderson2017efficientnaturallanguageresponse}, and apply Matryoshka Representation Learning \cite{kusupati2024matryoshka} to obtain nested embeddings that can be truncated for efficiency while retaining accuracy. The results show that domain-aware embeddings improve retrieval quality while preserving efficiency, laying the groundwork for richer marketplace search and enabling LLM-era experiences such as agentic assistants, real-time personalization, and natural-language exploratory search.

\section{Previous Work}

\subsection{Pre-Trained Query-Item Embeddings for Search Re-Ranking}

Before adopting pre-trained embeddings, Mercari's learning-to-rank (LTR) model used a task-specific embedding layer jointly trained with the ranking objective, which captured frequent lexical co-occurrences but generalized poorly to unseen words and semantic paraphrases. We then deployed ruri-small-v2 \cite{tsukagoshi2024rurijapanesegeneraltext} to encode queries and item titles; for each BM25-retrieved candidate pair, both cosine similarity and raw vectors served as continuous LTR features. To satisfy latency and memory constraints at scale, we compressed 768-dimensional embeddings to 32 dimensions via PCA, which preserved variance but reduced semantic precision for rare entities and nuanced query--item relations—motivating the domain-aware, truncation-robust approach presented next.

\subsection{Known Limitations}

While production performance on commonly used queries was strong, internal reviews and user feedback suggested unresolved gaps, such as brittle proper-noun handling in categories rich with brands/characters, clashes between general-domain priors and marketplace expectations, suboptimal term-importance calibration (generic modifiers outweighing core descriptors), and difficulty honoring nuanced constraints. To move from anecdote to evidence, we ran a focused qualitative assessment alongside offline log-based metrics and a public STS check, which guided the domain-aware fine-tuning described in the following sections.

\section{Methods}

\subsection{Training Setup}

\subsubsection{Loss Function and Dataset}

We fine-tune a text encoder with using Multiple Negatives Ranking (MNR) \cite{henderson2017efficientnaturallanguageresponse}. For a minibatch of positive query--title pairs and a similarity function, MNR treats all non-matching titles in the batch as in-batch negatives for each query, avoiding explicit hard-negative mining and sharpening separation among semantically close marketplace items. We train on 5 million positive query--title pairs randomly sampled from April--July 2025 marketplace logs: ``Query'' is the user's search query; ``Title'' is the title of an item \textit{purchased} within that query session.

\subsubsection{Query-Item Prefixes}

Following prior work on role-specific prefixes to model query--document asymmetry \cite{li2023generaltextembeddingsmultistage,wang-etal-2024-improving-text}, we attach role-specific \emph{Japanese} prefixes to inputs adhering to the base model convention \cite{tsukagoshi2024rurijapanesegeneraltext}. Translated into English, these prefixes are: a ``Query:'' prefix for user queries and a ``Passage:'' prefix for item titles; however, in production the actual prefix strings are written in Japanese. This lightweight conditioning allows us to use the same text encoder for both queries and items while explicitly modeling query--item asymmetry.

\subsubsection{Dimensionality Reduction}

For practicality and scalability, our target is 32-dimensional embeddings. To make representations truncation-robust, we adopt Matryoshka Representation Learning (MRL) \cite{kusupati2024matryoshka}, wrapping the base contrastive loss with nested objectives over the leading $d_1 > d_2 > \cdots > d_K$ dimensions and aggregating:
\[
\mathcal{L}_{\mathrm{MRL}} = \sum_{k=1}^{K} w_k \, \mathcal{L}_{\mathrm{base}}\!\big(q_{[:d_k]},\, t_{[:d_k]}\big).
\]
This encourages leading coordinates to remain informative so that truncation minimizes quality loss.

\subsection{Offline Evaluation Methods}

We evaluate from three angles: (i) manual qualitative assessment with domain stakeholders, (ii) offline evaluation on past search logs, and (iii) generalization on a public Japanese STS benchmark\footnote{\url{https://github.com/yahoojapan/JGLUE}} to check for catastrophic forgetting. \textit{Manual qualitative assessments:} we run a workshop-style review comparing top-$k$ retrievals from the baseline and fine-tuned model (32-D to mirror production), judged on proper nouns/disambiguation, in-domain context, term-importance alignment, and nuanced intent. \textit{Past search logs:} we construct a temporal hold-out split disjoint from training months. For each query, we retain the system-retrieved candidate set and user feedback events with graded relevance: \emph{purchase} $>$ \emph{like} $>$ \emph{comment} $>$ \emph{click} $>$ \emph{view}. \textit{Public Japanese STS benchmark:} to assess general-language retention in Japanese and detect catastrophic forgetting, we evaluate sentence-pair similarity on a public STS benchmark and report correlations between cosine similarity and gold labels (Pearson’s $r$ and Spearman’s $\rho$). We compare the base encoder against the fine-tuned model (using MNR+MRL) and measure robustness across truncated dimensions.

\section{Experiments}

\subsection{Manual Qualitative Assessments}

Side-by-side evaluations of retrieval quality for the base and fine-tuned models – both in 32 dimensions to assess performance  in the production environment –were carried out via a custom-built  web interface that issues a similarity search for a given query and displays the top-k results from both models in parallel, including item images and similarity scores. The cutoff k is user-adjustable (default k=10) to allow assessors to inspect different impression depths. As mentioned in the previous section, the reviewers judged each query’s results along the following axes: proper-noun understanding, in-domain user context, term importance alignment, and queries with complex nuance. Overall, the qualitative review indicates that the fine-tuned model outperforms the base model on proper-noun disambiguation, in-domain context, and term-importance alignment. In contrast, for queries with complex nuances, both models perform similarly, with neither showing a clear advantage. 

In the manual qualitative assessment, several representative patterns emerged. For the proper-noun query ``Switch 2'' (intent: the Nintendo Switch 2 console), the fine-tuned model surfaced more console listings and closely related peripherals, aligning better with the intended sense than the base model. For the ambiguous marketplace query ``coach,'' where domain context matters, the fine-tuned model reliably returned COACH-brand bags and fashion goods, whereas the base mixed brand items with general ``coaching'' or unrelated results. For term-importance sensitivity (e.g., ``ultra rare CHANEL belt,'' where the priority should be \emph{belt} $>$ \emph{CHANEL} $>$ \emph{rare}), the fine-tuned model emphasized belts and CHANEL accessories, while the base often over-weighted ``rare,'' retrieving unrelated items. Finally, for nuanced negation (e.g., ``something to wear for a wedding party, not too flashy''), both models struggled to consistently honor the ``not too flashy'' constraint. Collectively, these examples illustrate the fine-tuned embeddings' strengths in marketplace-specific semantics and proper-name disambiguation, alongside remaining challenges in handling complex negation.

\subsection{Quantitative Evaluation on Past Search Logs}

We report metrics at a cutoff of \(k=100\): \(\mathrm{nDCG@k}\), \(\mathrm{nDCG@k}\) (long), \(\mathrm{Precision@k}\), and \(\mathrm{Recall@k}\). We explored several values for \(k\): while small \(k\) (e.g., 5--10) aligns with UI slots for directly displayed results, our embeddings are used across multiple stages (candidate retrieval and re-ranking features), where a larger \(k\) better reflects downstream usage and yields more stable, generalizable estimates. Empirically, metric curves across \(k\in\{5,10,20,50,100\}\) stabilized at higher cutoffs; consistent with prior internal practice, we therefore adopt \(k=100\) for reporting. The test set contains 11{,}822 queries, of which 4{,}803 are \emph{long} (queries with \(\geq 10\) characters in Japanese) to probe cases requiring richer semantic understanding. For each query, we evaluate the top-100 retrieved items (total 1{,}182{,}200 query--item pairs). The full-dimension results are in Table~\ref{tab:full-d}. Relative to the strongest baseline (ruri-small-v2), the fine-tuned model improves \(\mathrm{nDCG@k}\) (0.198 \(\rightarrow\) 0.213; \(+7.6\%\)), \(\mathrm{nDCG@k}\) (long) (0.234 \(\rightarrow\) 0.245; \(+4.7\%\)), \(\mathrm{Precision@k}\) (0.015 \(\rightarrow\) 0.016; \(+6.7\%\)), and \(\mathrm{Recall@k}\) (0.613 \(\rightarrow\) 0.666; \(+8.6\%\)).

Under 32-D constraints, Matryoshka truncation plus domain fine-tuning yields substantial gains (Table~\ref{tab:trunc-32d}). Relative to the production PCA baseline, $\mathrm{nDCG@k}$ nearly doubles ($+97\%$), with large lifts for $\mathrm{nDCG@k}$ (long) ($+65\%$), $\mathrm{Precision@k}$ ($+114\%$), and $\mathrm{Recall@k}$ ($+123\%$). The improvements reflect both the weakness of PCA compression (0.198 $\rightarrow$ 0.099 at 32-D) and MRL's truncation-robust training.

\begin{table}[t]
\centering
\caption{Offline evaluation on past marketplace search logs with full-dimension embeddings. The fine-tuned model improves all metrics over the strongest baseline.}
\label{tab:full-d}
\footnotesize
\setlength{\tabcolsep}{3pt}
\renewcommand{\arraystretch}{0.95}
\begin{adjustbox}{max width=\linewidth}
\begin{tabular}{lrrrrr}
\toprule
\textbf{Model} & \textbf{\# dim} & \textbf{nDCG@k} & \textbf{nDCG@k (long)} & \textbf{Prec@k} & \textbf{Recall@k} \\
\midrule
Base            & 768 & 0.198 & 0.234 & 0.015 & 0.613 \\
\textbf{Fine-tuned} & \textbf{768} & \textbf{0.213} & \textbf{0.245} & \textbf{0.016} & \textbf{0.666} \\
\bottomrule
\end{tabular}
\end{adjustbox}
\end{table}

\begin{table}[t]
\centering
\caption{Offline evaluation on the same logs with 32-dimensional embeddings. At \(k=100\), Matryoshka truncation plus domain fine-tuning yields \(\approx 2\times\) nDCG@k versus production PCA and the best overall results.}
\label{tab:trunc-32d}
\footnotesize
\setlength{\tabcolsep}{3pt}
\renewcommand{\arraystretch}{0.95}
\begin{adjustbox}{max width=\linewidth}
\begin{tabular}{lrrrrr}
\toprule
\textbf{Model} & \textbf{\# dim} & \textbf{nDCG@k} & \textbf{nDCG@k (long)} & \textbf{Prec@k} & \textbf{Recall@k} \\
\midrule
Base            & 32 (PCA)   & 0.099 & 0.142 & 0.007 & 0.272 \\
\textbf{Fine-tuned} & \textbf{32 (MRL)} & \textbf{0.195} & \textbf{0.235} & \textbf{0.015} & \textbf{0.607} \\
\bottomrule
\end{tabular}
\end{adjustbox}
\end{table}

\subsection{Public Japanese STS benchmark}

We evaluate general-language similarity on JGLUE STS, reporting Pearson's $r$ and Spearman's $\rho$. At full dimensionality, the fine-tuned model is comparable to the base (Pearson $0.8648 \rightarrow 0.8393$; Spearman $0.8188 \rightarrow 0.7858$). Under truncation, we argue that the degradation is graceful (Table~\ref{tab:jsts}). This result indicates no catastrophic forgetting and that the FT model’s correlations on a general Japanese benchmark remain high, with only small, expected drops relative to the base encoder. The modest deltas are consistent with the intended domain shift, e.g., tokens like ``Coach'' aligning with the fashion brand rather than ``football manager'' in marketplace contexts, while overall semantic similarity remains strong enough to generalize beyond in-domain queries.

\begin{table}[t]
\centering
\caption{JGLUE Japanese STS (validation). Pearson $r$ and Spearman $\rho$ for base vs.\ fine-tuned models at 768d/256d/32d.}
\label{tab:jsts}
\footnotesize
\setlength{\tabcolsep}{4pt}
\renewcommand{\arraystretch}{0.95}
\begin{adjustbox}{max width=\linewidth}
\begin{tabular}{lrr}
\toprule
\textbf{Model (dims)} & \textbf{Pearson $r$} & \textbf{Spearman $\rho$} \\
\midrule
\textbf{Base (768d)}  & \textbf{0.8648} & \textbf{0.8188} \\
\emph{FT (768d)}      & \emph{0.8393}   & \emph{0.7858}   \\
FT (256d)             & 0.8360          & 0.7840          \\
FT (32d)              & 0.7934          & 0.7586          \\
\bottomrule
\end{tabular}
\end{adjustbox}
\end{table}

\subsection{Online A/B Test}

\begin{figure*}[t]
  \centering
  \includegraphics[width=\textwidth]{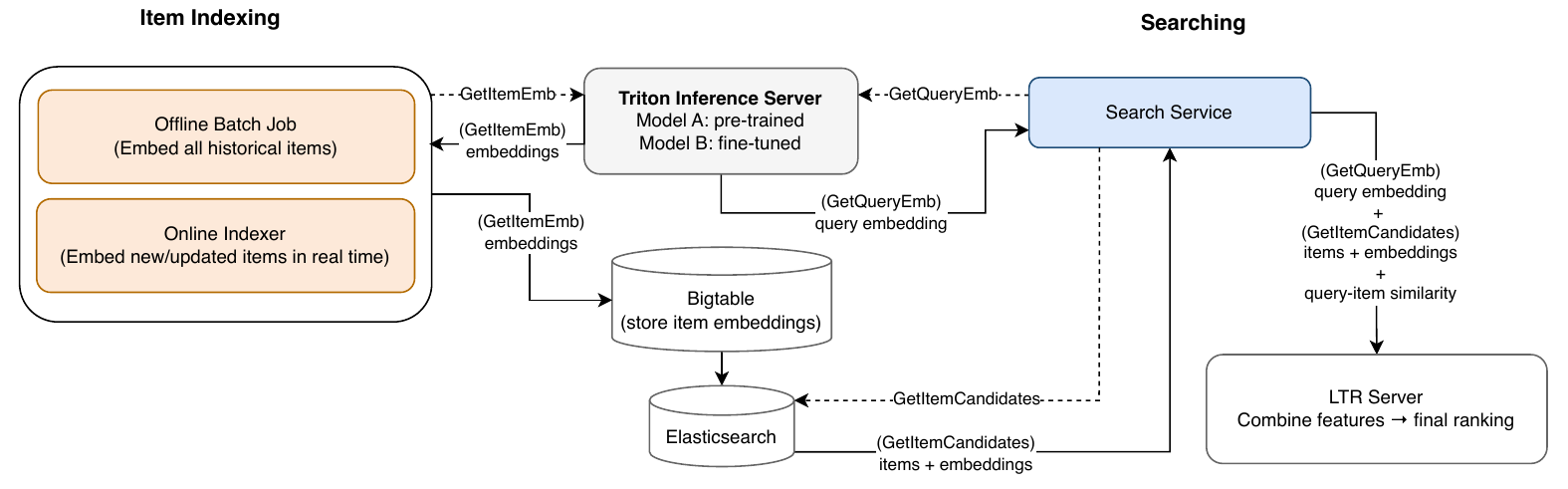}
  \caption{Production serving architecture for the A/B test. Control and Treatment differ only in the embedding model routed via Triton; all other components are identical. Solid lines indicate data flow, dashed lines indicate API calls between services.}
  \label{fig:serving-arch}
\end{figure*}

\subsubsection{Experiment Setup}
We conducted a seven-day online A/B test to evaluate the impact of fine-tuned, domain-aware text embeddings on real user search behavior.

\begin{itemize}
  \item \textbf{Control.} The existing production setup using \emph{pre-trained} query--item embeddings as continuous features to the learning-to-rank (LTR) model.
  \item \textbf{Treatment.} The same LTR configuration, but with \emph{fine-tuned, domain-aware} query--item embeddings replacing the pre-trained ones.
\end{itemize}

All other components---candidate retrieval, re-ranking, and serving infrastructure---were held fixed to isolate the effect of the embedding change. We monitored a broad set of online behavioral metrics to capture potential effects on user experience and business outcomes. The key metrics that showed statistically significant or practically meaningful differences are:

\begin{itemize}
  \item \textbf{Average Transactions per User (ATPU):} total transactions divided by active users (purchase frequency).
  \item \textbf{Average Revenue per User (ARPU):} gross merchandise value divided by active users (revenue contribution).
  \item \textbf{Average Order Value (AoV):} mean item price per transaction (purchase value).
  \item \textbf{Item Tap Rank:} position of the first clicked item in search results (ranking quality).
  \item \textbf{Average Impression via Search:} number of items viewed before engagement (search efficiency).
  \item \textbf{Buyer Conversion Rate (BCR):} the ratio of users who make a purchase to the total number of users.
  \item \textbf{Unique Item View (UIV):} the number of unique items viewed.
\end{itemize}

\subsubsection{Serving and Deployment}
The experiment was deployed on the production search infrastructure (Figure~\ref{fig:serving-arch}). The main components are as follow:

\begin{itemize}
    \item \textit{Embedding model serving.} Query and item encoders produce vector representations via a shared Triton inference server. The server hosts both the pre-trained and fine-tuned variants; routing configuration selects the appropriate model per bucket, enabling seamless switching between Control and Treatment without other system changes.
    \item \textit{Item indexing.} A one-time batch inference pass generated embeddings for existing item titles. These embeddings were written to the Bigtable feature store and propagated to Elasticsearch (ES) prior to the experiment. For newly listed or updated items, embeddings are generated synchronously via Triton at write time and stored in the same feature store.
    \item \textit{Querying.} At query time, the search service obtains a query embedding from Triton and retrieves candidate items---with their stored embeddings---from ES. The query and item features, including both embeddings and their cosine similarity, are then forwarded to the LTR server for final re-ranking.
\end{itemize}

\subsubsection{Result and Practical Implications}

As a first step, we conducted a one-week A/B test comparing the fine-tuned embedding model (treatment) against the baseline (control) on live marketplace traffic. As shown in Table~\ref{tab:abtest}, the treatment group delivered a statistically significant uplift in monetization, with ARPU increasing by $+0.92\%$ ($p<0.05$) driven by AoV $+0.91\%$ ($p<0.05$), while ATPU remained non-statistically significant. We also observed improved search efficiency: users interacted with relevant items earlier (Item Tap Rank $-0.65\%$, $p<0.05$) and required fewer exposures to reach satisfactory results (Average Impression via Search $-0.64\%$, $p<0.05$).

Additionally, we then evaluated hybrid retrieval---blending lexical matching with semantic candidates retrieved by vector similarity---in a two-week online A/B experiment. The treatment group, which integrates semantic search results when similarity score $>0.90$, produced consistent gains. To focus on the product surface directly governed by hybrid retrieval, we report metrics measured on the best-match (BM) search module, i.e., the portion of the search engine results page (SERP) that directly applies the hybrid retrieval logic when displaying results (other SERP surfaces may show items through different mechanisms). Under the treatment group, buyer conversion rate (BCR) increased by $+0.88\%$ and ATPU increased by $+0.96\%$, alongside higher engagement/coverage (unique item view (UIV) $+1.48\%$ and impressions $+2.25\%$; Table~\ref{tab:hybrid_abtest}). The impact was especially pronounced on sparse-result SERPs: hybrid retrieval recovered $60.2\%$ of would-be zero-hit SERPs and $66.1\%$ of would-be low-hit SERPs, with substantial downstream lifts on low-hit SERPs (UIV $+12.0\%$, BCR $+6.95\%$).

Taken together, these experiments provide consistent online evidence that domain-adapted semantic representations and hybrid retrieval can improve relevance and revenue while reducing search effort, with outsized benefits on low/zero-hit queries.

\begin{table}[t]
\centering
\caption{Online A/B test results comparing Control (base embedding model) and Treatment (fine-tuned embedding model). Treatment shows statistically significant improvements in revenue per user and search engagement efficiency.}
\label{tab:abtest}
\footnotesize
\setlength{\tabcolsep}{3pt}
\renewcommand{\arraystretch}{0.95}
\begin{adjustbox}{max width=\linewidth}
\begin{tabular}{l r}
\toprule
\textbf{Metric} & \textbf{Lift (Treatment / Control)} \\
\midrule
Average Transaction per User (ATPU) & Non stat-sig \\
Average Revenue per User (ARPU) & +0.92\% ($p<0.05$) \\
Average Order Price (AoV) & +0.91\% ($p<0.05$) \\
Item Tap Rank & $-0.65\%$ ($p<0.05$) \\
Average Impression via Search & $-0.64\%$ ($p<0.05$) \\
\bottomrule
\end{tabular}
\end{adjustbox}
\end{table}

\begin{table}[t]
\centering
\caption{Hybrid retrieval online experiment. Reported lifts are for the best-match (BM) search module, the SERP component that directly uses the hybrid retrieval logic. Low-hit SERP metrics are reported for the treatment group only.}
\label{tab:hybrid_abtest}
\footnotesize
\setlength{\tabcolsep}{3pt}
\renewcommand{\arraystretch}{0.95}
\begin{adjustbox}{max width=\linewidth}
\begin{tabular}{l r}
\toprule
\textbf{Metric (BM)} & \textbf{Lift (Treatment / Control)} \\
\midrule
Daily BCR & +0.88\% \\
Daily ATPU & +0.96\% \\
Daily impression count & +2.25\% \\
Daily UIV & +1.48\% \\
\midrule
\multicolumn{2}{l}{\textit{Low-hit SERP metrics}} \\
Zero-hit recovery rate & 60.2\% \\
Low-hit recovery rate & 66.1\% \\
UIV on low-hit SERPs & +12.0\% \\
BCR on low-hit SERPs & +6.95\% \\
\bottomrule
\end{tabular}
\end{adjustbox}
\end{table}

\section{Conclusion}

We investigated the impact of domain-aware text embeddings for Japanese C2C search at Mercari, combining MNR-based text encoder fine-tuning on purchase-driven pairs with Matryoshka representations for truncation-robust efficiency. Across offline log replay, public STS, qualitative review, and an initial online A/B test, our approach consistently improved retrieval quality. In production, the A/B test showed ARPU $+0.92\%$ ($p<0.05$) driven by AoV $+0.91\%$. Users also reached satisfactory results faster, as indicated by Item Tap Rank $-0.65\%$ and Impressions per engagement $-0.64\%$ (both $p<0.05$). Replacing PCA with MRL at 32-D closed much of the compression gap, nearly doubling nDCG@k relative to the production baseline, while preserving broad semantic competence on STS.

While the model substantially improves proper-noun disambiguation, in-domain semantics, and coverage (Recall@k), nuanced intents (e.g., negation or soft style constraints) remain challenging. Limitations include reliance on implicit negatives from in-batch sampling, potential bias from purchase-only positives, and limited exposure to multimodal or code-switched inputs. The reported A/B test is the first step in operationalizing these embeddings; going forward, we will integrate them more broadly across the search and discovery stack, including multimodal search, hybrid retrieval (dense + sparse), real-time personalization, and agentic conversational search workflows.

\section{Acknowledgment}

We are grateful to Kentaro Takiguchi and Pathompong Yupensuk for their involvement in the model training and evaluation process; their expertise and familiarity with Mercari’s search ecosystem informed several key decisions throughout this work. We also acknowledge Aman Kumar Singh and Julio Christian Young for their efforts in deploying the model to production for the A/B test, which allowed us to assess and validate the model’s impact in real-world conditions. Finally, we thank the members of the Search, Recommendation, ML Platform, and AI/LLM teams for their contributions, which supported the completion of this project and facilitated the model’s deployment across additional marketplace domains, thereby extending its practical impact and providing valuable feedback to help guide future research directions.

\bibliography{aaai2026}

\end{document}